# LucidRaster: GPU Software Rasterizer for Exact Order-Independent Transparency

Krzysztof Jakubowski

## Abstract

Transparency rendering is problematic and can be considered an open problem in real-time graphics. There are many different algorithms currently available, but handling complex scenes and achieving accurate, glitch-free results is still costly.

This paper describes LucidRaster: a software rasterizer running on a GPU which allows for efficient exact rendering of complex transparent scenes. It uses a new two-stage sorting technique and sample accumulation method. On average it's faster than high-quality OIT approximations and only about 3x slower than hardware alpha blending. It can be very efficient especially when rendering scenes with high triangle density or high depth complexity.

LucidRaster is mainly implemented in C++ & GLSL shaders and is using Vulkan API. Full source code is available on GitHub: https://github.com/nadult/lucid

## 1. Introduction

Rendering 3D scenes with transparent surfaces can be difficult. Graphics hardware for a long time had support for blending transparent surfaces, but these surfaces had to be sorted first. In general, order-independent transparency (OIT) techniques allow for rendering transparent scenes without this requirement. An exact OIT algorithm has to identify all the fragments which contribute to a given pixel, sort them and blend them in the correct order. Good examples of exact OIT algorithms are a-buffer [Carpenter 1984], and depth peeling [Everitt 2001]. Advanced algorithms which utilize backwards memory allocation [Knowles 2013] and improved sorting techniques [Knowles 2014] can provide big performance improvements.

Still, exact OIT techniques are about an order of magnitude slower than unsorted hardware alpha-blending. Approximate OIT techniques fill the performance gap between the two. Weighted-blended OIT (WBOIT) [McGuire 2013] is only about 20% slower than hardware alpha-blending. Moment-based OIT (MBOIT) [Münstermann 2018] is significantly costlier than WBOIT but provides a much better approximation.

Two surveys cover a wide range of OIT algorithms: "A Survey of Multifragment Rendering" [Vasilakis 2020] and "Exploring and Expanding the Continuum of OIT Algorithms" [Wyman 2016].

With each next generation, GPUs become more and more suitable for general-purpose computations. Current GPUs already allow for efficient implementation of software rasterizers.

There are several interesting examples available. CudaRaster [Laine 2011] and cuRE [Kenzel 2018] are general-purpose rasterizers. FreePipe [Liu 2010] similarly to LucidRaster has a focus on multi-fragment effects. All three of these rasterizers are several times slower than hardware with CudaRaster being the fastest. Nanite [Karis 2021] uses a simple software micro-poly rasterizer which can be up to 3x faster than a general-purpose GPU rasterizer.

Most of the OIT algorithms are built around a hardware rasterizer, which is used for the accumulation and / or composition of fragments. There were successful attempts at using compute shaders for sprite rendering [Köhler 2016], but in general, the core computation in a typical OIT algorithm happens in a fragment shader. This is a major restriction which limits the potential efficiency of OIT algorithms. LucidRaster is an attempt to remove this restriction by implementing a software rasterizer with custom pipeline tailored for transparency rendering.

LucidRaster performance is evaluated by comparing it with a simple renderer which performs hardware-based unsorted alpha blending. Moment-based OIT paper [Peters 2018] contains a similar comparison which allows to roughly compare LucidRaster with several different OIT algorithms:

| Algorithm | Parameters | Relative running time |
|---|---:|---:|
| Hardware alpha blending | | 1.00 |
| Weighted-blended OIT | | 1.17 |
| Multi-layer alpha blending (MLAB) | 2 layers | 2.78 |
| Multi-layer alpha blending | 4 layers | 4.75 |
| Moment based OIT | 4 power moments, 80 bits | 2.64 |
| Moment based OIT | 6 power moments, 112 bits | 3.13 |
| Moment based OIT | 4 trigonometric moments, 144 bits | 5.00 |
| LucidRaster | | 3.30 |

The scores for WBOIT, MLAB and MBOIT are averages, relative to hardware alpha blending, taken from MBOIT paper. The score for LucidRaster is an average computed from SW/HW values for all scenes and all tested GPUs. Because the algorithms were tested in different environments, this table can provide only a very rough idea about the performance differences between other OIT algorithms and LucidRaster.

## 2. Pipeline description

LucidRaster is built around several core ideas:

- Sort-middle [Molnar 1994] pipeline where primitives are first distributed into small evenly-sized bins. This allows for sorting and merging of all fragments in a given bin in shared memory, which can save a lot of framebuffer bandwidth.
- Two-stage sorting: depth-sorting blocks of fragments (8x4 or 8x8) coupled with per-pixel fixed-size fragment sorter.
- Taking quads as input instead of triangles. This allows to save bandwidth and reduces processing time in the initial pipeline stages.

LucidRaster pipeline is comprised of three stages:

- Setup stage: input primitives are tested for visibility. For each visible primitive, shading and rasterization data are computed and stored.
- Binning stage: the screen is divided into 32x32-pixel bins and for each bin a list of overlapping primitives is created.
- Rasterization stage: bins are rasterized one by one: triangles in each bin are sorted, samples are generated, blended together and written to the framebuffer.

LucidRaster pipeline is quite similar to cuda-raster [Laine 2011]. The paper describing this rasterizer is also a very good introduction into software rasterization on GPUs.

## Stage 1: quad setup

Essentially setup stage takes a stream of input quads, it culls those which do not contribute to the final image and pre-compute some data for those which are visible. Pre-computed data will be used in later pipeline stages.

Input quads are divided into chunks of up to 1024 primitives which have the same material. Each chunk is processed by a different workgroup, every one having 1024 threads. Quad chunks are processed in two phases with a compacting step happening between them. The first phase culls invisible quads, which are then removed during compaction. The second phase does the pre-computation and storage.

**Phase 1: processing input quads**

Each thread handles one input quad. It loads its vertex indices and if both tris which comprise a quad are degenerate, then the whole quad is culled. Vertex positions are loaded and if backface culling is enabled, each triangle is tested (in world space) whether it is back- or front-facing the camera and if both triangles are back-facing then the quad is culled. Next, quad vertices are transformed to NDC space and vertices are tested if they are on the visible side of the clipping planes. If all vertices are on the outer side of at least one of the clipping planes then the quad is culled.

Because LucidRaster uses 3D rasterization [Davidovic 2012], full clipping is not required. This simplifies the setup stage, but screen-space AABBs still have to be computed. For that Jim Blinn's 'Calculating Screen Coverage' algorithm [Blinn 1996] is used. With AABBs computed, quads which fall between the rasterizer's sample positions can be detected and culled ([Laine 2011] p 5.1). Finally, quads which weren't culled up until this point are considered visible. Those quads are classified as either small or large based on the number of bins which they overlap. If it's more than 4 then quad is classified as large.

Once all visible quads in a chunk are identified, they are compacted and space for them is allocated in global memory. Compacting improves performance of the second phase in cases where large percentage of randomly distributed quads are culled. Shared memory is used to transfer visible quads' data from the first phase to the second phase to avoid recomputation. For example, AABBs encoded in single uint are transferred this way.

**Phase 2: writing primitives data**

In this phase for each visible quad, shading & rasterization data are computed and stored in global memory. Different pre-computed quad attributes are put into different storages. This division is based on the way the data will be accessed later in the pipeline: attributes which are read together are stored together. Some attributes are computed and stored per-triangle, others per-quad.

Per-triangle attributes are put in 4 different storages with the following contents:

- Triangle normals (encoded as uint: X10Y10Z10)
- Edge functions for barycentric coordinates (vec4[2])
- Edge function for depth, instance flags and instance id (encoded in uvec4)
- 3D edge functions for scanline rasterizer, screen space AABB in the Y axis (encoded in uvec4[2])

Per-quad attributes are put in 4 different storages with the following contents:

- bin-resolution AABBs for the whole quad and per-triangle culling info (as uint: 28 bits for AABB and 2 bits for culling info)
- vertex colors (as uvec4, each color encoded in R8G8B8A8)
- vertex normals (as uvec4, each normal encoded in X10Y10Z10)
- vertex texture coordinates (as vec4[2])

Per-quad vertex attributes are optional and are only stored if the quad's material actually needs them. They require almost no computation and are basically copied from vertex buffers, but when grouped together in per-quad storage they improve cache efficiency in the final rasterization phase and in effect improve performance of the whole pipeline.

## Stage 2: quad binning

Binning stage takes a flat stream of visible quads divided in two groups: small and large. At this stage, the main task is to generate a list of overlapping primitives for each bin. The secondary task is to categorize bins based on primitive density. Quad binning happens in 3 phases, each involving a separate compute shader dispatch.

**Phase 1: counting per-bin primitives**

In this phase only as many workgroups are spawned as required to saturate all GPU cores. Workgroups work in a persistent threading [Gupta 2012] fashion: each in a loop tries to acquire and process a batch of primitives until all primitives are processed. Workgroups have to be large enough so that they could fit an array of 32-bit per-bin primitive counters in shared memory. With a resolution limit of 2560x2048 and with 32x32 bins there can be at most 5120 bins. Depending on the platform, 512 or 1024 threads per workgroup should be enough to hold this array in shared memory without negatively affecting occupancy.

The main goal in this phase is to count, for each bin, the number of overlapped primitives. Primitives are processed in batches. Because there are two categories of primitives (large and small), there are also two types of batches: quad-batches which contain small quads and tri-batches with large triangles. Because small quads require very little processing, it's best to process more quads at once to minimize per-batch overhead. Tri-batches on the other hand usually take a lot more time to process, because rasterization is involved and the primitive surface area is larger.

Additionally, in a typical scene, there aren't that many of them. So tri-batches are only large enough so that each thread in a workgroup has a single triangle to work on.

Quad-batches and tri-batches are processed separately: tri-batch processing starts only after all quad-batches have been processed. Each workgroup processes some number of batches of each type by counting (in a shared memory array) per-bin instances of overlapped primitives. To count small quads, only AABBs are used. It's not enough information to count accurately, but because small quads can overlap at most 2x2 bins, in most cases the estimate will be accurate. In the worst case, 1 bin out of 4 will be marked incorrectly as overlapped which will negligibly affect the rasterizer stage performance. For large triangles, rasterization is used. Once there are no more batches of a given type to process, two things happen. First, all per-bin counters are atomically accumulated from all workgroups into a single array in global memory (BIN_QUAD_COUNTS for quads and BIN_TRI_COUNTS for triangles). Secondly, each workgroup saves to global memory a list of processed batches (a single number is sufficient to identify a batch) and per-bin counters array. In a way, it is a snapshot, which will allow to continue working on the same set of batches in the 3rd phase.

**Phase 2: bin categorization & offset computation**

Once the first phase is finished, BIN_QUAD_COUNTS and BIN_TRI_COUNTS arrays will contain primitive overlap counts for each bin. In this phase, these arrays are used to perform two tasks, each with a single workgroup. The first task is to compute offsets for primitive lists for each bin by applying prefix sum to per-bin primitive counts. These offsets will be used in the next phase when storing indices of per-bin primitives. The second task is to categorize bins based on the number of overlapping triangles (quads are counted as 2 triangles). Currently, there are 3 categories: empty, low (for bins with less than 1024 triangles) and high (for bins with at least 1024 triangles).

**Phase 3: writing per-bin primitive lists**

The third phase is in many ways similar to the first. It runs the same number of workgroups. Each workgroup continues work from the first phase workgroup with the same id. It will generate per-bin primitive lists for the same primitives which were processed in the first phase (by using a previously saved batch list). Previously saved per-workgroup per-bin counters are used to allocate space in a global per-bin offsets array for primitive indices for all bins touched by selected batches. After allocation, in shared memory, in an array similar as in the first phase, global per-bin offsets are kept.

Quad-batches and tri-batches are processed separately, just like in the first phase. Processing primitives is also similar, but instead of counting, for each overlapped bin, an offset is computed by atomically incrementing a per-bin counter in shared memory and writing an appropriate primitive index to global memory. Large triangles may cause load-balancing problems, because surface area of different triangles may vary wildly. To deal with that, a simple balancing scheme is implemented: threads within a [subgroup](subgroup) try to equally divide between them to-be-written indices on a row level. This scheme is simple and can provide a solid performance boost on some scenes (phase 3 up to 2.5X faster).

It might be concerning that per-workgroup batch lists are inherited from phase 1 to phase 3. But in practice, it doesn't cause any load balancing problems across workgroups, because the amount of work which each workgroup has to perform in phase 1 is proportional to the amount of work in phase 3.

**Triangle rasterization**

The triangle scanline rasterization algorithm which is used in LucidRaster is similar to micropoly software rasterizer used in Nanite [Karis 2021] with the main difference that 3D rasterization [Davidovic 2012] is also used. Basically, for a given triangle it iterates over all rows of its screen-space AABB. In each row instead of testing each pixel, it computes (in constant time) an interval of overlapped pixels. 3D edge functions with parameters precomputed during the setup stage are used.

There are two versions of the scanline algorithm: one which works on pixels (which is used in the final rasterization stage) and another which works on 32x32-pixel bins (this one is used in binning stage). The bin version is also optimized with a simple 'trivial reject test' [Abrash 2009].

## Stage 3: bin rasterization

During the bin rasterization stage, 32x32-pixel bins are further divided into smaller regions: 32x8-pixel block-rows, 8x8-pixel blocks and 8x4-pixel half-blocks. All of these are fixed on a grid: block-rows on 1x4 grid (4 total), blocks on 4x4 grid (16 total) and half-blocks on 4x8 grid (32 total). Per-triangle pixel coverage information can be generated for all of those. Let's name the data structures that hold per-triangle pixel coverage information: Tri-block-row for block-rows, tri-block for blocks and tri-half-block for half-blocks.

LucidRaster handles bins with low and high densities of triangles with two different compute shaders: low-rasterizer and high-rasterizer. The low-density version besides the limit on the number of triangles per bin also has limits on the number of fragments and triangles per block and per block-row. Those limits allow for more efficient processing but can be too constraining for some, albeit very small, percentage of bins. Such bins are detected in low-rasterizer and are propagated to high-rasterizer (which has much weaker constraints). Because of that high-rasterizer workgroups are dispatched only after the low-rasterizer has finished. What follows is a description of the low-rasterizer; the differences between the two versions are described afterwards.

Similarly to the binning stage, persistent threads are used in this stage as well. Each workgroup in a loop tries to acquire and rasterize a bin until all bins are processed. Again, only as many workgroups are spawned as required to saturate all GPU cores. Low-rasterizer uses workgroups containing 256 threads each.

Bin rasterization happens in three phases:

- Phase 1: Block-row generation: arrays of tri-block-rows are generated.
- Phase 2: Half-block extraction: depth-sorted tri-half-blocks are extracted from tri-block-rows.
- Phase 3: Shading & blending: tri-half-blocks are shaded, blended together and written to the output framebuffer.

**Phase 1: tri-block-row generation**

Phase 1 begins with processing all per-bin small quads and large triangles. Quads are further divided into triangles, and triangles are processed directly. For each triangle, scanline algorithm is used to generate intervals of covered pixels for each row within 32x8-pixel block-rows. Per-triangle scanline data includes minimum & maximum Y coordinates, which allows the scanline algorithm to iterate only over block-rows with not-empty intervals. Tri-block-row pixel coverage information can be efficiently encoded in less than 128 bits: (5 + 5) * 8 = 80 bits for per-row pixel intervals, 4 bits for block column coverage and 24 bits for visible triangle index. This information is saved in a per-workgroup scratch buffer stored in global memory. Tri-block-rows belonging to different block-rows are stored separately in 4 different arrays. Because of the limit on the number of triangles per block-row in low-rasterizer (1024), the size of a per-workgroup scratch buffer can be kept relatively low. Shared memory atomics are used for counting tri-block-rows.

**Phase 2: half-block extraction**

The goal in this phase is to generate, for all half-blocks within the current bin, a depth-sorted list of tri-half-blocks.

Subgroups process all blocks within a current bin one after another. Subgroups work independently, each working on a single block at the time. Each subgroup will generate tri-blocks, sort them and divide them into two arrays of tri-half-blocks in a scratch buffer in global memory.

Careful use of shared memory is the key to efficiency in this phase. Each subgroup has at its disposal 8 uints per thread (256 uints total for 32 threads). In the low-rasterizer, the upper limit on the number of triangles per block is 256, which allows to perform most of the processing (especially sorting) without constantly writing/reading to/from global memory.

Subgroups start by first identifying which tri-block-rows overlap the current block. Block column coverage bits are used for that. Each subgroup generates in shared memory a list of tri-block-row indices. The easiest way to do it (and also quite efficient) is to simply iterate over all tri-block-rows and for those which overlap the current block, use an atomic counter to compute an index and count them at the same time.

To generate a sorted tri-block list, depth values are needed as well. So, subgroups iterate over tri-block-row index lists and compute first tri-block sample centroids by using tri-block-row data and after that depth values with per-triangle edge functions. Depth values together with tri-block index (which is identical to tri-block-row index) are encoded into a single 32-bit uint and stored in shared memory. 22 high bits are used for depth, 10 low bits are used for tri-block-index. Those values are then sorted with bitonic sort optimized with subgroup shuffles [Demouth 2013]. Tri-blocks are sorted front to back.

At the end of the phase each subgroup iterates over sorted tri-block indices, generates two arrays of tri-half-blocks and stores them in scratch buffer in global memory. Tri-half-blocks contain:

- Pixel coverage information ((3 + 3) * 4 = 24 bits).
- Visible triangle index (24 bits).
- Fragment count prefix sum (only 12 lower bits are needed). This value is used in the next phase to quickly compute an index for each sample.

**Phase 3: shading & blending**

Once the rasterizer gets to phase 3, data required to efficiently render samples in half-blocks will be available in a scratch buffer in global memory. Now subgroups will independently shade & blend each half-block. LucidRaster currently works only on GPUs which have subgroups containing 32 or 64 threads. In the case of 64-thread subgroup, two half-blocks are processed at the same time by a single subgroup.

All samples in a given half-block are ordered. Samples coming from different triangles are ordered in the same way as tri-half-blocks and as for the samples coming from the same triangle, they are ordered row-by-row, with top-left sample being first and bottom-right being last. Samples are additionally grouped in segments, where each segment can contain at most 256 samples. All segments except the last one will be full and contain 256 samples.

Subgroups in a loop process samples in segments, one segment at a time. Each thread within a subgroup is responsible for a single pixel: it keeps the current RGBA color (initially fully transparent black) and depth filter context. This context contains a constant size (currently 3) of depth-color pairs. Depth filter allows for fixing invalid depth order of samples which may happen when tri-half-blocks sample depths overlap. It cannot fix all depth-order issues, but it handles the vast majority of cases. Increasing depth-filter size allows for solving more complex ordering issues, but increases register pressure and decreases overall performance.

Each thread in a subgroup takes a single sample from a segment and shades it. After shading those samples are immediately blended with previously blended samples. Sample color & depth has to be transferred to the appropriate thread based on the sample position. It can be done efficiently with bitwise operations and subgroup shuffles: after shading each thread atomically writes a single bit to a 32x32-bit array in shared memory. Bit position encodes sample position within half-block and thread-id. Afterwards, each thread iterates in-order over all samples which overlap the pixel position assigned to the current thread. Those samples are then blended with those which were processed previously.

Once all samples are processed, threads write the colors to the framebuffer. The workgroup can then proceed to the next bin.

**Sample shading**

In LucidRaster simple fragment shader is used to demonstrate the basic capabilities of the rasterizer. Shader supports vertex normals & colors, single texture per triangle and performs simple lighting calculation. To support mip-mapping, texture coordinate gradients are required. Instead of using quad-shading, analytic derivatives are computed. This allows for more efficient processing of tiny triangles. To support multiple textures in a given scene, texture atlases are used.

**Sample blending**

Samples, before being blended into the final result, first have to pass through the depth-filter. This filter is basically a fixed-size priority queue, in which samples are ordered by depth. Whenever a new sample is added to the stack, one of the samples falls out and is blended into the target result.

By increasing the depth-filter size by 1, pixels with incorrectly ordered samples can also be detected. Those pixels could be registered and afterwards processed by a slower, conservative algorithm. In most scenes less than 0.1% of pixels are invalid. In the most problematic scenes, it's between 0.5% - 1%. Those errors can be prevented by fixing the geometry (for example in white-oak and san-miguel scenes foliage is especially problematic, but intersections in foliage geometry are not required to achieve a satisfactory result) or by increasing depth-filter size. For example number of invalid pixels in white-oak scene can be decreased from 1% to 0.08% by increasing the depth-filter size from 3 to 8. This slows down the whole rendering performance by 1.5%.

Blending is performed front to back. This allows for a simple, but efficient optimization. When accumulated alpha approaches 1.0, then further samples can be discarded, because their contribution will be minimal. In LucidRaster this optimization can be enabled with 'alpha_threshold' flag. When enabled, shading & blending stops once all pixels within a half-block have accumulated at least 1 - 1 / 128 alpha value.

**High-rasterizer**

High-rasterizer is very similar to the low-rasterizer, but it has to be able to process a lot more triangles per-bin than the low-rasterizer. The main differences are the following:

- Increased workgroup size to 1024. This increases available shared memory per-bin and also improves caching efficiency.
- Per-workgroup scratch buffer is much bigger. Currently, 768 KB are used for the scratch buffer. This will fit 128K tri-block-rows (16K per block-row) and 128K tri-half-blocks (4K per half-block).
- In phase 2, tri-half-blocks are generated and sorted directly. The amount of space in shared memory required to generate and sort tri-half-blocks may be much bigger than in the low-rasterizer. Because of that, shared memory is allocated dynamically for this task and the number of subgroups working on a given half-block may vary as well.

# 3. Results

**Test scenes**

12 different scenes were used during the evaluation of LucidRaster. To simplify the renderer all textures are packed into texture atlases. Each scene can use at most 2 atlases: one opaque and one transparent. Block compression is used: BC1 for opaque and BC3 for transparent atlases.

Triangles are grouped into quads using a greedy maximal independent set algorithm [Halldorsson 1994]. The algorithm is performed on a graph where quads are treated as nodes and triangles (which are shared by quads) as edges. In most scenes, this gives good results with on average about 5% of quads being degenerate.

In the next table, for each scene, the following information is provided: number of triangles & quads, percentage of visible quads for a given view (**vis**), percentage of degenerate quads (**deg**), whether the backface culling is turned on (**BC**) or not.

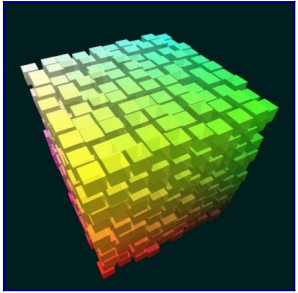 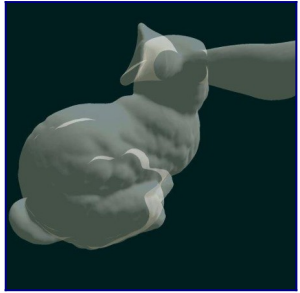 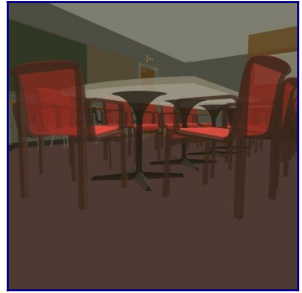 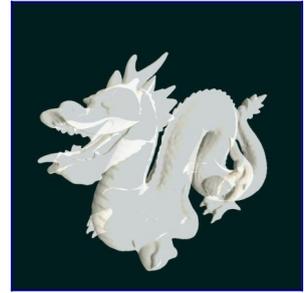

Boxes: 12K tris, 6K quads, 100% vis, 0% deg

Bunny: 144K tris, 72K quads, 40% vis, 0.3% deg, BC

Conference: 331K tris, 167K quads, 6% vis, 1.9% deg, BC

Dragon: 871K tris, 438K quads, 42% vis 0.9% deg, BC

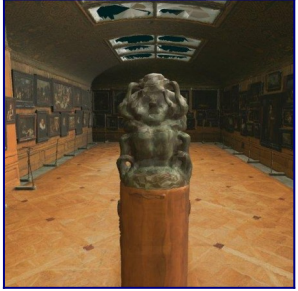 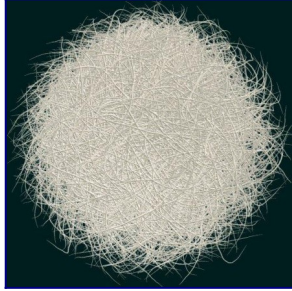 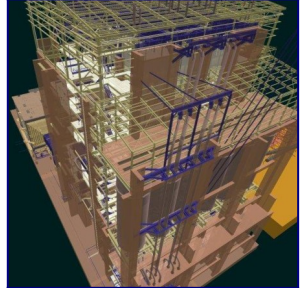 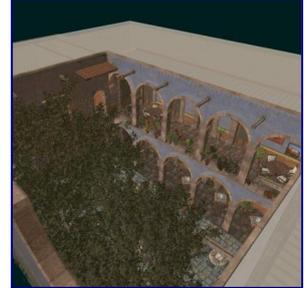

Gallery: 999K tris, 513K quads, 60% vis, 5.2% deg

Hairball: (2880K tris, 1495K quads, 99% vis, 7.4% deg

Powerplant: 12759K tris, 6402K quads, 15% vis, 0.7% deg, BC

San miguel: 5617K tris, 3069K quads, 34% vis, 17% deg

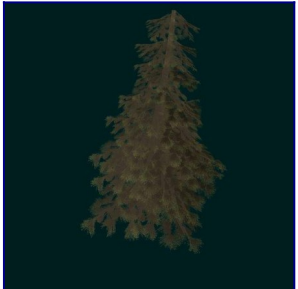 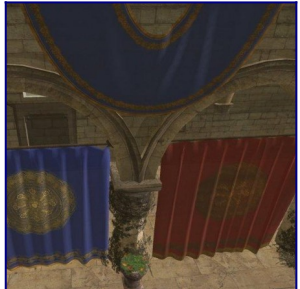 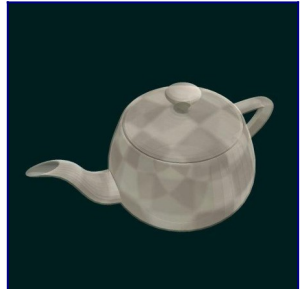 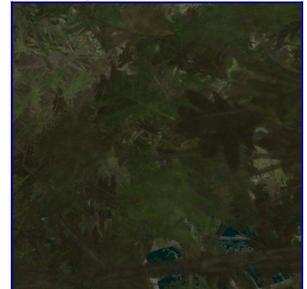

Scrub pine: 527 tris, 289 quads, 100% vis, 18% deg

Sponza: 262K tris, 132K quads, 18% vis, 1.5% deg

Teapot: 16K tris, 8K quads 100% vis, 0.1% deg

White oak: 37K tris, 19K quads, 53% vis, 10% deg

## Performance results

LucidRaster performance was compared with a simple renderer which renders all triangles using hardware with alpha blending on and without any kind of sorting. **SW**, **HW** represent, respectively, LucidRaster and simple hardware-based renderer running times in microseconds. **S/THB** is the average number of samples divided by the number of tri-half-blocks.

The fragment shader is almost identical in both renderers, with the main differences in vertex attribute retrieval and interpolation. Fragment shader code is available on GitHub: https://github.com/nadult/lucid/blob/main/data/shaders/simple_material.glsl

Measurements were gathered in 'mixed' mode where in a single frame both simple and lucid renderer were running one after another. Timings shown in tables are averages of samples gathered over several seconds for each scene.

Measurements on AMD Radeon RX 6700 XT were done on Linux, because it seems that the latest AMD Vulkan drivers for Linux have better subgroup support. Specifically, windows drivers didn't support VK_EXT_subgroup_size_control and some shaders which used subgroup instructions were working several times slower than on Linux.

| NVidia GTX 3050 TI, Windows 10, 2560x1330 | | | | | LucidRaster (SW) stages | | | |
|---|---|---|---|---|---|---|---|---|
| Scene | Samples | S/THB | SW | HW | SW/HW | setup | binning | low_raster | hi_raster |
| Boxes | 12.11M | 17.91 | 1707 | 401 | 4.26 | 24 | 53 | 1609 | 4 |
| Bunny | 0.74M | 3.33 | 350 | 78 | 4.49 | 53 | 38 | 234 | 4 |
| Conference | 5.31M | 16.75 | 1158 | 226 | 5.12 | 58 | 76 | 1002 | 4 |
| Dragon | 0.67M | 1.25 | 837 | 248 | 3.38 | 328 | 51 | 302 | 136 |
| Gallery | 4.89M | 3.07 | 2621 | 678 | 3.87 | 771 | 78 | 1427 | 324 |
| Hairball | 22.00M | 2.89 | 11518 | 3564 | 3.23 | 1756 | 244 | 355 | 9146 |
| Powerplant | 18.03M | 3.60 | 10121 | 4761 | 2.13 | 2720 | 346 | 838 | 6201 |
| San miguel | 21.27M | 6.57 | 9234 | 3667 | 2.52 | 2391 | 216 | 2261 | 4349 |
| Scrub pine | 3.26M | 24.44 | 575 | 193 | 2.98 | 15 | 61 | 474 | 4 |
| Sponza | 24.00M | 19.80 | 3910 | 1690 | 2.31 | 98 | 130 | 3658 | 3 |
| Teapot | 1.71M | 8.53 | 373 | 111 | 3.36 | 23 | 40 | 285 | 4 |
| White oak | 122.00M | 26.95 | 13005 | 6049 | 2.15 | 37 | 189 | 12759 | 3 |
| | | | | Averages: | 3.32 | 14% | 5% | 60% | 18% |
| AMD Radeon RX 6700 XT, Ubuntu 22.04, 1920x1031 | | | | | LucidRaster (SW) stages | | | |
| Scene | Samples | S/HB | SW | HW | SW/HW | setup | binning | low_raster | hi_raster |
| Boxes | 7.28M | 15.80 | 546 | 202 | 2.70 | 10 | 17 | 508 | 6 |
| Bunny | 0.44M | 2.48 | 131 | 32 | 4.09 | 17 | 19 | 84 | 5 |
| Conference | 3.11M | 14.77 | 370 | 101 | 3.66 | 19 | 44 | 296 | 5 |
| Dragon | 0.40M | 1.15 | 272 | 95 | 2.86 | 88 | 25 | 68 | 85 |
| Gallery | 2.86M | 2.48 | 907 | 224 | 4.05 | 186 | 31 | 453 | 232 |
| Hairball | 13.22M | 2.38 | 4388 | 1084 | 4.05 | 991 | 116 | 152 | 3123 |
| Powerplant | 10.83M | 3.27 | 3532 | 1533 | 2.30 | 1260 | 144 | 237 | 1885 |
| San miguel | 12.59M | 5.65 | 3396 | 1066 | 3.19 | 1127 | 97 | 653 | 1513 |
| Scrub pine | 1.96M | 22.90 | 242 | 57 | 4.25 | 6 | 24 | 200 | 6 |
| Sponza | 13.97M | 17.66 | 1194 | 375 | 3.18 | 33 | 55 | 1052 | 47 |
| Teapot | 1.03M | 6.91 | 147 | 50 | 2.94 | 11 | 18 | 107 | 5 |
| White oak | 71.41M | 25.68 | 3500 | 1624 | 2.16 | 11 | 80 | 3397 | 7 |
| | | | | Averages: | 3.29 | 15% | 7% | 57% | 20% |

**Alpha threshold**

When the 'alpha threshold' mode is enabled, shading stops once enough alpha value is accumulated in a given pixel. Rendering of some scenes can be much faster when this flag is used, especially in cases when lots of samples are hidden beneath an almost opaque layer of samples. The following speedups were achieved:

| NVidia GTX 3050 TI, Windows 10, 2560x1330 | | alpha_threshold off (default) | | alpha_threshold on | |
|---|---|---|---|---|---|
| Scene | Scene opacity | HW | SW | SW/HW | SW | SW/HW |
| Boxes | 0.5 | 401 | 1707 | 4.26 | 1288 | 3.21 |
| Boxes | 0.8 | 401 | 1707 | 4.26 | 1169 | 2.92 |
| Hairball | 0.5 | 3564 | 11518 | 3.23 | 9784 | 2.75 |
| Powerplant | 0.5 | 4761 | 10121 | 2.13 | 9445 | 1.98 |
| White oak | 0.5 | 6049 | 13005 | 2.15 | 13129 | 2.17 |
| White oak | 0.8 | 6049 | 13005 | 2.15 | 11070 | 1.83 |

In other scenes the gains were negligible.

**Fragment sorting errors**

Two-stage sorting technique fails in some cases. Those cases are rare and can be detected and dealt with properly. On average the number of invalid pixels is very small (0.17%, 0.02% with DF=8), so even if they were handled with a very slow shader, it shouldn't affect overall performance noticeably. LucidRaster currently doesn't handle those cases, but it will show invalid pixels when the 'visualize_errors' flag is enabled.

The number of invalid pixels depends on the geometrical complexity of nearby surface intersections in the scene and the size of the depth filter. What follows is a table with error statistics for all scenes for different depth filter sizes.

|  | Invalid pixels (total) | | | Invalid pixels (percentage) | | | Rendering time compared to DF=3 | |
|---:|---:|---:|---:|---:|---:|---:|---:|---:|
| **Scene** | **DF=3** | **DF=8** | **DF=12** | **DF=3** | **DF=8** | **DF=12** | **DF=8** | **DF=12** |
| Boxes | 3 | 0 | 0 | <0.01% | | | 102% | 115% |
| Bunny | 1 | 0 | 0 | <0.01% | | | 101% | 124% |
| Conference | 19 | 0 | 0 | <0.01% | | | 104% | 127% |
| Dragon | 0 | 0 | 0 | | | | 99% | 108% |
| Gallery | 70 | 1 | 0 | <0.01% | <0.01% | | 101% | 121% |
| Hairball | 14100 | 1530 | 660 | 0.41% | 0.04% | 0.02% | 100% | 107% |
| Powerplant | 1800 | 15 | 3 | 0.05% | <0.01% | <0.01% | 100% | 106% |
| San miguel | 17112 | 4513 | 2828 | 0.50% | 0.13% | 0.08% | 103% | 112% |
| Scrub pine | 190 | 0 | 0 | 0.01% | | | 102% | 121% |
| Sponza | 290 | 4 | 1 | 0.01% | <0.01% | <0.01% | 103% | 128% |
| Teapot | 9 | 0 | 0 | <0.01% | | | 103% | 124% |
| White oak | 34780 | 2850 | 905 | 1.02% | 0.08% | 0.03% | 103% | 116% |
| | | | Averages: | 0.17% | 0.02% | 0.01% | 102% | 117% |

In the next table, a visualization of errors in the most problematic areas of 3 different scenes (Hairball, San miguel and White oak) is shown.

| Depth filter size = 3 (default) | Depth filter size = 8 | Depth filter size = 12 |

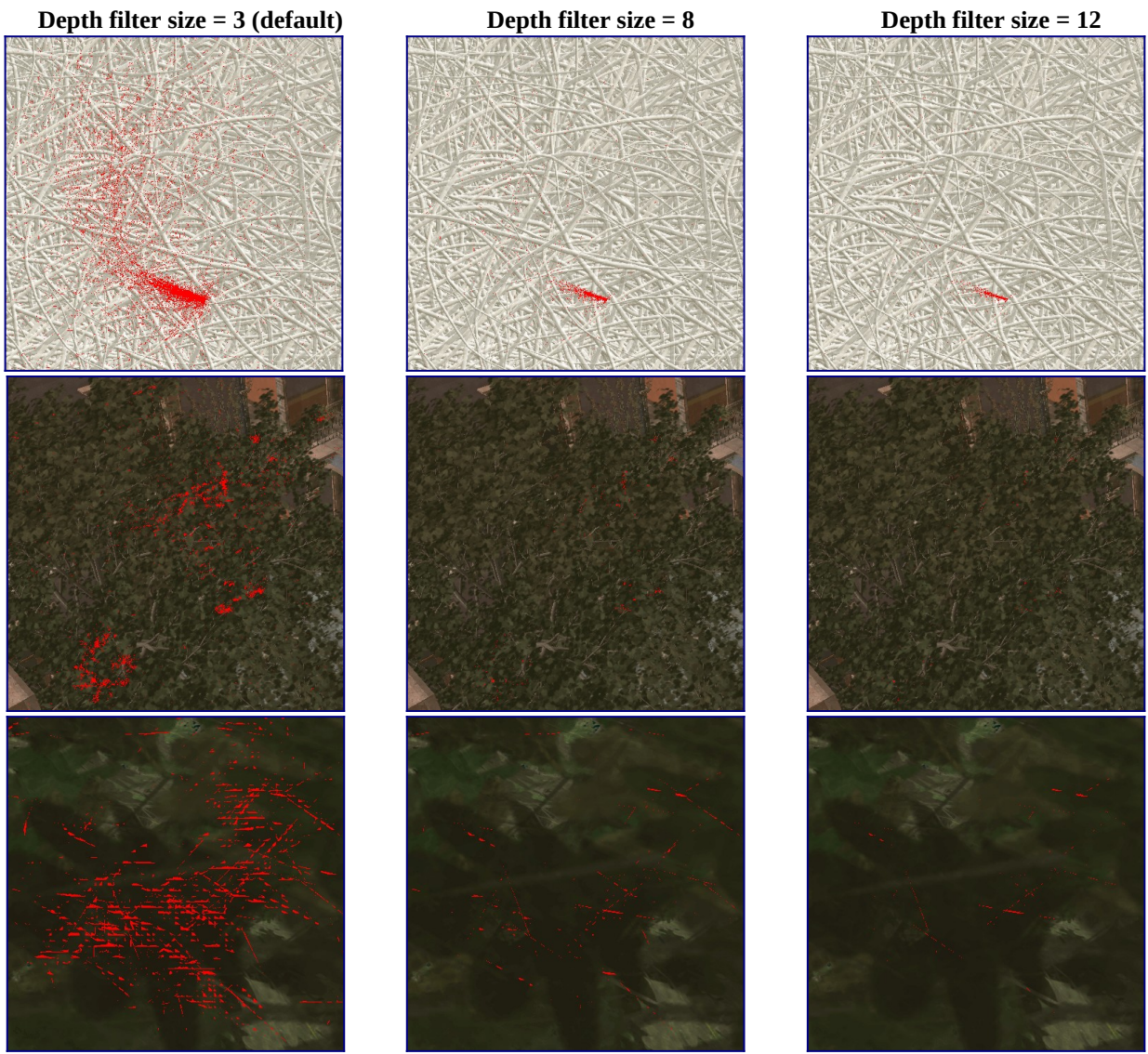

# 4. Future work

Work on LucidRaster was completely self-funded. One of the main goals was to create a minimum prototype which demonstrates the usefulness of this rendering method. Because of that LucidRaster uses only a simple shading model, all textures are packed into atlases and even the rendering resolution is quite limited (to 2560x2048). Additionally, LucidRaster uses a lot of GPU memory: about 1200MB to hold all the scratch, triangle & quad buffers, to be able to render scenes with millions of triangles. Many of these restrictions are not necessary, but they were added to simplify implementation. It should be possible to greatly reduce memory usage with dynamic allocation because most of the time most of the GPU storages are mostly empty.

LucidRaster currently has no anti-aliasing support, but it shouldn't be hard to extend it with 4x or 16x multisample anti-aliasing (MSAA). MSAA would mostly affect the first and second phases of the raster stage. More space would be needed to store triangle sample ranges (~2x more for 4x MSAA and ~4x more for 16x MSAA). High-quality MSAA should be achievable just by modifying the blending phase and without a big increase in the number of samples (alpha values for samples would be affected by the number of MSAA samples).

Although the pipeline is very well optimized, there are still areas for improvement. For example, currently, LucidRaster uses 32x32 bins, but it could be beneficial to use 64x64 bins in areas with low density of triangles and 16x16 in high-density areas. It should also be possible to run high-rasterizers and low-rasterizers at the same time, which hopefully would increase parallelism.

# 5. Vulkan glossary

**Workgroup**: A collection of compute shader threads which execute the same program in parallel. Threads within a workgroup can share resources and synchronise their execution.

**Subgroup**: A collection of compute shader threads which can effectively communicate and synchronize.

**Global memory**: Off-chip GPU memory (DRAM). Accessible in Vulkan's compute shaders mainly through buffers, images and textures. It's global in the sense that it can be shared between different workgroups and pipeline stages.

**Shared memory**: Fast, small on-chip GPU memory shared across all threads within a workgroup.

More information about Vulkan and compute shaders can be found in:

- Vulkan 1.3 specification:
  https://registry.khronos.org/vulkan/specs/1.3-extensions/html/vkspec.html
- Compute shader 101 glossary:
  https://github.com/googlefonts/compute-shader-101/blob/main/docs/glossary.md
- Vulkan subgroup tutorial: https://www.khronos.org/blog/vulkan-subgroup-tutorial